\documentclass{aastex}
\usepackage{spr-astr-addons}
\usepackage{url}

\RequirePackage{color}

\begin{document}

\title{Quasinormal Modes and Hawking radiation of a Reissner-Nordstr\"om black hole surrounded by quintessence}
\slugcomment{Not to appear in Nonlearned J., 45.}
%% Running heads
\shorttitle{QNMs and Hawking radiation of R-N black hole}
\shortauthors{Mahamat S. et al.}

\author{Mahamat Saleh\altaffilmark{1}} \and \author{Bouetou Bouetou Thomas
        \altaffilmark{2}}
 \and
\author{Timoleon Crepin Kofane}
\affil{Department of Physics, Faculty of Science, University of
Yaounde I, P.O. Box 812, Cameroon}

\email{mahsaleh2000@yahoo.fr}

\altaffiltext{1}{Department of Physics, Higher Teachers' Training
College, University of Maroua, P.O. Box 55 Maroua, Cameroon}
\altaffiltext{2}{Ecole Nationale Sup$\acute{e}$rieure Polytechnique,
University of Yaounde I, P.O. Box 8390, Cameroon}

\begin{abstract}
We investigate quasinormal modes(QNMs) and Hawking radiation of a Reissner-Nordstr\"om black hole surrounded by quintessence. The
Wentzel-Kramers-Brillouin (WKB) method is used to evaluate the QNMs and the rate of radiation. The results show that due to the interaction of
the quintessence with the background metric, the QNMs of the black hole damp more slowly when increasing the density of quintessence and the
black hole radiates at slower rate.
\end{abstract}


%\section*{}
%\label{sec:intro}
\section{Introduction}
\label{intro} The investigation of the general properties of black
holes has attracted considerable interest. It is well known that
when perturbing a black hole, it undergoes oscillations which are
characterized by some complex eigenvalues, with negative imaginary
parts, of the wave equations called quasinormal frequencies. Their
real parts represent the oscillation frequencies, while the
imaginary ones determine the damping rates of modes. QNMs of black
holes have been an intriguing subject of discussion for the last
few decades(\cite{Siyer, KokSchutz, Seidel, Kok, vckl, yuzg1,
yuzg2, mah}). On the other hand, Hawking radiation from black
holes is one of the most striking effects that is known, or at
least widely agreed to arise from combination of quantum mechanics
and general relativity. As one of the most important achievements
of quantum field theory in curved spacetime, the discovery of
hawking radiation supported these ideas availably which showed
that a classical black hole could radiate thermal spectrum of
particles. Hawking and Ellis(\cite{haw}) evoke the possible origin
of radiation to be a black body radiation left over from a hot
early stage of the universe, the result of superposition of a very
large number of very distant unsolved discrete sources or
intergalactic grains which thermalize other forms of radiations.
The Hawking radiation for the Reissner-Nordstr\"om black hole is
widely studied in literature(\cite{jiang, yu, zhangxu, zhao1}).

There has been growing observational evidence showing that our
universe is accelerated expanding driven by a yet unknown dark
energy. Recent results from the cosmic microwave background(CMB)
combined with supernova of type Ia, large-scale structure (cosmic
shear) and galaxy cluster abundances show that our universe is
dominated by a mysterious dark energy with negative
pressure($\sim70\%$) and contains  cold dark matter with
negligible pressure($\sim25\%$), the ordinary baryonic matter
makes up $5\%$(\cite{muk}). Dark energy can be studied by its
influence on the expansion of the universe as well as on the
growth of the large scale structure. Cosmological models with a
dark energy fluid with an equation of state parameter $\omega_q$
close to $-1$ are favored by combining recent CMB, supernova and
baryon acoustic oscillations data, suggesting that the Hubble
expansion accelerates in the current cosmic epoch. There are
several types of models of  dark energy such as the cosmological
constant(\cite{rol}), quintessence(\cite{var, kis, shu}),
phantom(\cite{mart}), k-essence(\cite{yang}), and
quintom(\cite{zong, jun}) models. For quintessence, the equation
of state parameter is in the range of $-1\leq\omega_{q}\leq-1/3$.
Recently, \cite{kis} considered Einstein's field equation
surrounded by quintessential matter and obtained a new solution
dependent on state parameter $\omega_{q}$ of the quintessence.
\cite{hod} investigated the late-time evolution of charged
gravitational collapse and decay of charged scalar hair.
\cite{konop2} investigated the decay of charged scalar field in
the Reissner-Nordstr\"om black hole background.  In this paper, a
Reissner-Nordstr\"om black hole surrounded by quintessence is
considered to investigate QNMs and Hawking radiation including the
influence of the quintessence on them.

The paper is organized as follows. In section ~\ref{sec:1}, we derive the wave equation of a scalar perturbation in the Reissner-Nordstr\"om
background surrounded by quintessence. In section ~\ref{sec:2}, we evaluate the QN frequencies of the scalar perturbation by using the third
order WKB approximation method. In section ~\ref{sec:3}, Hawking radiation is investigated. The last section is devoted to a summary and
conclusion.

\section{Scalar field perturbation}
\label{sec:1} For the Reissner-Nordstr\"om black hole, the metric
is given by:
\begin{equation}\label{1}
ds^{2}=-\big(1-\frac{2M}{r}+\frac{Q^{2}}{r^{2}}\big)dt^{2}+(1-\frac{2M}{r}+\frac{Q^{2}}{r^{2}})^{-1}dr^{2}+r^{2}d\Omega^{2},
\end{equation}
where $M$ is the black hole mass and $Q$, the charge of the black
hole.

Due to the interaction of the quintessence with the spacetime, the
background metric transforms to (\cite{mah})

\begin{equation}\label{2}
ds^{2}=-f(r)dt^{2}+f(r)^{-1}dr^{2}-r^{2}d\Omega^{2},
\end{equation}
 with $f(r)=\big(1-\frac{2M}{r}+\frac{Q^{2}}{r^{2}}-\frac{c}{r^{3\omega_{q}+1}}\big)$, $\omega_{q}$ is the quintessential state parameter, $c$
the normalization factor related to the density of quintessence,
$\rho_{q}=-\frac{c}{2}\frac{3\omega_{q}}{r^{3(\omega_{q}+1)}}$.

\begin{figure}[h]
  \includegraphics[width=0.45\textwidth]{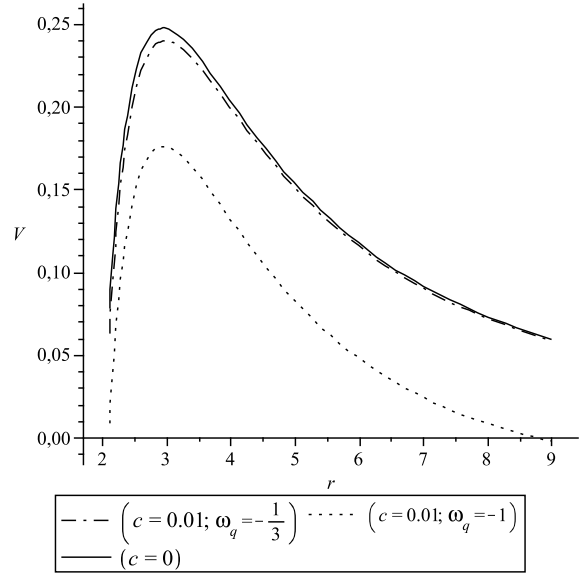}
\caption{Black hole's effective potential without
quintessence($c=0$) and with quintessence($c=0.01$; $\omega_q=-1$
and $\omega_q=-1/3$)}
\label{fig:1}       % Give a unique label
\end{figure}

Using the tortoise coordinate $r_{*}$ defined by
$dr_{*}=\big(1-\frac{2M}{r}+\frac{Q^{2}}{r^{2}}-\frac{c}{r^{3\omega_{q}+1}}\big)^{-1}dr$,
the metric can be rewritten as:
\begin{equation}\label{3}
ds^{2}=f(r)(-dt^{2}+dr^{2}_{*})-r^{2}(d\theta^{2}+sin^{2}\theta
d\varphi^{2}).
\end{equation}
We consider the evolution of massless scalar perturbation. The
wave equation for the complex scalar field is given by
(\cite{haw}):
\begin{equation}\label{4}
    \phi_{;\mu\nu}g^{\mu\nu}-\frac{k^{2}}{\hbar^{2}}\phi=0,
\end{equation}
where $k$ and $\hbar$ are constants. $m=\frac{k}{\hbar}$ represents the mass of the scalar field.\\
We represent the scalar field into spherical harmonics
$\phi=\sum_{l,m}^{} \psi^{l}_{m}( r)e^{-i\omega t}Y^{m}_{l}(\theta,
\varphi)/r$ and after some algebra  the equation of motion takes the
form:
\begin{equation}\label{5}
    \psi_{,r_{*}r_{*}}+(\omega^{2}-V)\psi=0,
\end{equation}
where the black hole potential
\begin{equation}\label{6}
    V=f(r)\bigg[\frac{l(l+1)}{r^{2}}+\frac{2M}{r^{3}}-\frac{2Q^{2}}{r^{4}}+\frac{c(3\omega_{q}+1)}{r^{3\omega_{q}+3}}+\frac{k^{2}}{\hbar^{2}}\bigg].
\end{equation}
is represented in figure~\ref{fig:1}. Since we considered a massless
scalar field, $k=0$.

Through this figure, we can see that the non-quintessential
potential is higher than the quintessential ones, and the height
of the potential decreases with increasing $\omega_q$.
\section{Quasinormal modes}
\label{sec:2} The wave equation (~\ref{5}) can be rewritten as:
\begin{equation}\label{7}
    \frac{d^{2}\psi}{dr^{2}_{*}}+Q(r)\psi=0,
\end{equation}
where $Q(r)=\omega^2-V$. For a black hole, the QN frequencies
correspond to solution of perturbation equation which satisfy the
boundary conditions appropriate for purely ingoing waves at the
horizon and purely outgoing waves at infinity. Incoming and
outgoing waves correspond to the radial solution proportional to
$e^{-i\omega r_{*}}$ and $e^{i\omega r_{*}}$, respectively. Only a
discrete set of complex frequencies satisfies these conditions.

To evaluate the QN frequencies, we applied the third order WKB
approximation method derived by Schutz, Will(\cite{SW})and
Iyer(\cite{SIW}) to the above equation and these QN frequencies
are given by(\cite{yuzg2})
\begin{equation}\label{8}
    \omega^{2}=[V_{0}+(-2V''_{0})^{1/2}\tilde{\Lambda}]-i(n+\frac{1}{2})(-2V''_{0})^{1/2}[1+\tilde{\Omega}],
\end{equation}
where \begin{eqnarray}
      \nonumber
      \begin{tabular}{lll}
      $\tilde{\Lambda}$ &=& $\frac{1}{(-2V''_{0})^{1/2}}\bigg\{\frac{1}{8}(\frac{V^{(4)}_{0}}{V''_{0}})(\frac{1}{4}+\alpha^{2})$\\
      &&$-\frac{1}{288}(\frac{V'''_{0}}{V''_{0}})^{2}(7+60\alpha^{2})\bigg\}$ \\
     $ \tilde{\Omega}$
     &=&$\frac{1}{-2V''_{0}}\bigg\{\frac{5}{6912}(\frac{V'''_{0}}{V''_{0}})^{4}(77+188\alpha^{2})$\\
     && $-\frac{1}{384}(\frac{V'''^{2}_{0}V^{(4)}_{0}}{V''^{3}_{0}})(51+100\alpha^{2})$\\
    &&
     $-\frac{1}{288}(\frac{V^{(6)}_{0}}{V''_{0}})(5+4\alpha^{2})+\frac{1}{288}(\frac{V'''_{0}V^{(5)}_{0}}{V''^{2}_{0}})(19+28\alpha^{2})$\\
     &&$+\frac{1}{2304}(\frac{V^{(4)}_{0}}{V''_{0}})^{2}(67+68\alpha^{2})\bigg\}$,
     \end{tabular}
      \end{eqnarray}
      $\alpha=n+\frac{1}{2}$, and
      $V^{(n)}_{0}=\frac{d^{n}V}{dr^{n}_{*}}|_{r_{*}=r_{*}(r_{p})}$.

      Using equation(~\ref{8}), we calculated numerically the QN frequencies of the scalar field perturbation for $M=1,
      Q=0.1$  without quintessence and with  quintessence.  The results are shown in the following tables where $l$ is the
harmonic angular index, $n$ is the overtone number, $\omega$ is
the complex QN frequency and $\omega_{q}$ is the state parameter
of the quintessence.

\begin{table*}
%table caption is above the table
\small \caption{Quasinormal modes of scalar field perturbation
without quintessence }
\label{tab:1}       % Give a unique label
\begin{tabular}{llllll}
\tableline
l & $\omega(n=0)$ & $\omega(n=1)$  & $\omega(n=2)$   & \multicolumn{1}{c}{$\omega(n=3)$} & $\omega(n=4)$ \\
\tableline
2 & 0.4840 - 0.0969i & 0.4640 - 0.2960i & 0.4326 - 0.5037i & 0.3936 - 0.7162i &  \\
3 & 0.6763 - 0.0966i & 0.6616 - 0.2925i & 0.6361 - 0.4944i & 0.6035 - 0.7014i & 0.5651 - 0.9116i \\
4 & 0.8688 - 0.0964i & 0.8572 - 0.2911i & 0.8360 - 0.4898i & 0.8080 - 0.6930i & 0.7747 - 0.8998i \\
5 & 1.0613 - 0.0964i & 1.0518 - 0.2903i & 1.0338 - 0.4872i & 1.0093 - 0.6878i & 0.9798 - 0.8918i \\
\tableline
\end{tabular}
%\end{table*}
%\begin{table*}
\small \caption{Quasinormal modes of scalar field perturbation
with quintessence for $c=0.01$ and $k=0$}
\label{tab:2}       % Give a unique label
\begin{tabular}{lllllll}
\tableline
l & $\omega_q$ & $\omega(n=0)$ & $\omega(n=1)$  & $\omega(n=2)$   & \multicolumn{1}{c}{$\omega(n=3)$} & $\omega(n=4)$ \\
\tableline
 & -1 & 0.4091 - 0.0840i & 0.3974 - 0.2545i & 0.3779 - 0.4286i & 0.3519 - 0.6049i \\
 & -5/6 & 0.4425 - 0.0880i & 0.4261 - 0.2682i & 0.3998 - 0.4551i & 0.3666 -
 0.6459i \\
2 & -2/3 & 0.4609 - 0.0911i & 0.4425 - 0.2783i & 0.4133 - 0.4734i
&
  0.3772 - 0.6729i\\
  & -1/2 & 0.4710 - 0.0934i & 0.4519 - 0.2854i & 0.4216 - 0.4856i & 0.3842 -
  0.6904i \\
    & -1/3 & 0.4767 - 0.0949i & 0.4572 - 0.2900i & 0.4264 - 0.4935i & 0.3883 - 0.7017i \\
\tableline
 & -1 & 0.5747 - 0.0832i & 0.5658 - 0.2508i & 0.5497 - 0.4211i & 0.5281 - 0.5938i & 0.5015 - 0.7680i \\
 & -5/6 & 0.6199 - 0.0875i & 0.6077 - 0.2648i & 0.5862 - 0.4467i & 0.5585 - 0.6328i & 0.5255 - 0.8215i \\
3  & -2/3 & 0.6447 - 0.0908i & 0.6311 - 0.2750i & 0.6074 - 0.4646i & 0.5772 - 0.6590i & 0.5415 - 0.8563i \\
  & -1/2 & 0.6584 - 0.0931i & 0.6443 - 0.2820i & 0.6198 - 0.4766i & 0.5885 - 0.6762i & 0.5516 - 0.8788i \\
  & -1/3 & 0.6661 - 0.0946i & 0.6517 - 0.2866i & 0.6268 - 0.4844i & 0.5949 - 0.6872i & 0.5574 - 0.8932i \\
\tableline
  & -1 & 0.7400 - 0.0828i & 0.7329 - 0.2493i & 0.7196 - 0.4176i & 0.7012 - 0.5880i & 0.6786 - 0.7600i \\
  & -5/6 & 0.7972 - 0.0873i & 0.7875 - 0.2633i & 0.7697 - 0.4426i & 0.7459 - 0.6255i & 0.7174 - 0.8112i \\
 4 & -2/3 & 0.8285 - 0.0907i & 0.8178 - 0.2736i & 0.7982 - 0.4602i & 0.7722 - 0.6511i & 0.7413 - 0.8452i \\
 & -1/2 & 0.8459 - 0.0930i & 0.8348 - 0.2807i & 0.8145 - 0.4722i & 0.7876 - 0.6681i & 0.7556 - 0.8674i \\
  & -1/3 & 0.8557 - 0.0945i & 0.8444 - 0.2852i & 0.8237 - 0.4800i & 0.7963 - 0.6791i & 0.7638 - 0.8816i \\
\tableline
 & -1 & 0.9051 - 0.0827i & 0.8992 - 0.2485i & 0.8879 - 0.4157i & 0.8721 - 0.5846i & 0.8523 - 0.7550i \\
 & -5/6 & 0.9745 - 0.0872i & 0.9664 - 0.2626i & 0.9514 - 0.4403i & 0.9306 - 0.6210i & 0.9054 - 0.8044i \\
5 & -2/3 & 1.0124 - 0.0906i & 1.0035 - 0.2729i & 0.9870 - 0.4578i & 0.9642 - 0.6462i & 0.9368 - 0.8377i \\
 & -1/2 & 1.0335 - 0.0930i & 1.0243 - 0.2800i & 1.0071 - 0.4698i & 0.9836 - 0.6631i & 0.9553 - 0.8598i \\
  & -1/3 & 1.0454 - 0.0945i & 1.0360 - 0.2845i & 1.0186 - 0.4775i & 0.9946 - 0.6740i & 0.9658 -
  0.8739i\\
\tableline
\end{tabular}
\end{table*}

\begin{figure}[h!]
  \includegraphics[width=0.45\textwidth]{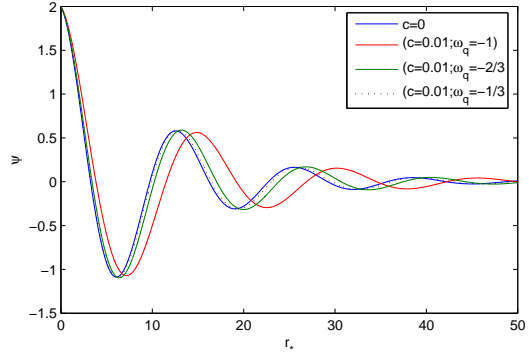}
\caption{Behavior of the scalar perturbation for some frequencies
with($l=2$, $n=0$)}
\label{fig:2}       % Give a unique label
\end{figure}

We then plot the behavior of the scalar perturbation for some
frequencies. The results are shown in figure \ref{fig:2}.

\begin{figure}[h!]
  \includegraphics[width=0.47\textwidth]{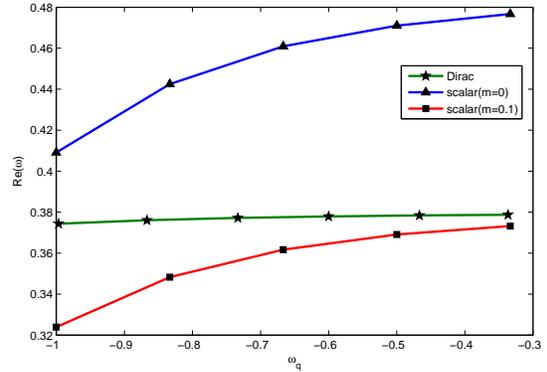}
\caption{Comparison of the real parts of $\omega$ when varying
$\omega_q$ ($l=2$, $n=0$)}
\label{fig:3}       % Give a unique label
\end{figure}

\begin{figure}[h!]
  \includegraphics[width=0.45\textwidth]{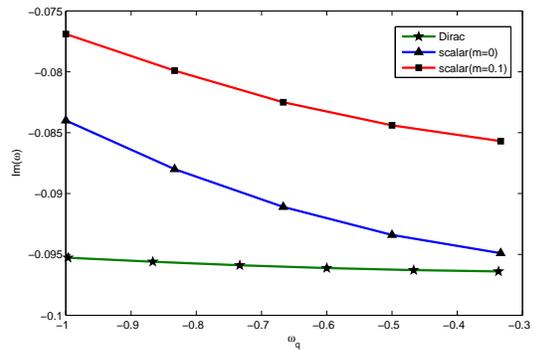}
\caption{Comparison of the imaginary parts of $\omega$ when varying
$\omega_q$ ($l=2$, $n=0$)}
\label{fig:4}       % Give a unique label
\end{figure}

The QNMs of the Reissner-Nordstr\"om black hole surrounded by
quintessence were investigated for the Dirac field by \cite{yuzg3}
and for the charged massive scalar field by \cite{nijo}. Comparing
these results, we pointed out that the massless scalar field
oscillates more rapidly than the Dirac field which oscillates more
rapidly than the massive scalar field. But in term of damping, Dirac
field damps more rapidly than scalar fields. On the other hand, the
massless scalar field damps more rapidly than the massive one. When
increasing the state parameter of quintessence $\omega_q$, the real
part and the absolute value of the imaginary part of $\omega$ are
increasing for the scalar fields but their variation is negligible
for the Dirac field. Moreover, the rate of variation for the
massless scalar field is higher than that of the massive one(see
Figures \ref{fig:3} and \ref{fig:4}).

\section{Hawking radiation}
\label{sec:3} Let's rescale the time coordinate into
Eddington-Finkelstein coordinate(\cite{zhangxu})

$t=T\pm r_*,$

where the signs $+$ and $-$ represent ingoing and outgoing
particles, respectively. The tortoise coordinate $r_*$ is defined
as
\begin{equation}\label{9}
    \frac{dr_*}{dr}=f(r)^{-1}.
\end{equation}
In the following, the study is restricted to the outgoing particle
radiated from the black hole horizon.

The background metric can then transform to
\begin{equation}\label{10}
    ds^{2}=-f(r)dT^{2}+2dTdr+r^{2}(d\theta^{2}+sin^{2}\theta
    d\varphi^{2}).
\end{equation}

The metric obtained is a Vaidya-Bonner like metric(\cite{zhen})
and can represent a Vaidya-Bonner black hole surrounded by
quintessence.

The apparent horizon of this metric is given by the following
equation
\begin{equation}\label{11}
    1-\frac{2M}{r}+\frac{Q^{2}}{r^{2}}-\frac{c}{r^{3\omega_{q}+1}}=0.
\end{equation}
In the absence of quintessence($c=0$), this equation gives two
solutions $r_\pm=M\pm\sqrt{M^{2}-Q^{2}}$.

Actually, the normalization factor related to the density of
quintessence, $c$, is smaller than $0.001$(\cite{yuzg1}).
Therefore, the contribution to the metric background due to the
presence of quintessence can be treated as a perturbation.

We regard the quintessence as perturbation leading to a small
modification of the horizon radii and we put then the
quintessential horizon radii $R_\pm$ in the form
\begin{equation}\label{12}
    R_{\pm}=r_{\pm}+\epsilon_{\pm}.
\end{equation}

Substituting this expression into Eq.(11), we obtain
\begin{equation}\label{13}
\begin{tabular}{ll}
    $1-\frac{2M}{r_{\pm}}\Big(1-\frac{\epsilon_{\pm}}{r_{\pm}}\Big)+\frac{Q^{2}}{r^{2}_{\pm}}\Big(1-\frac{2\epsilon_{\pm}}{r_{\pm}}\Big)$&\\
    $-cr^{-3\omega_{q}-1}_{\pm}\Big(1-(1+3\omega_{q})\frac{\epsilon_{\pm}}{r_{\pm}}\Big)$&=0
    \end{tabular}
\end{equation}

which gives us
\begin{equation}\label{14}
    \epsilon_{\pm}\simeq\frac{cr^{1-3\omega_{q}}_{\pm}}{r_{\pm}-r_{\mp}},
\end{equation}
in first approximation.

The radial null geodesic is given by
\begin{equation}\label{15}
    \dot{r}=\frac{dr}{dT}=\frac{1}{2}\Big(1-\frac{2M}{r}+\frac{Q^{2}}{r^{2}}-\frac{c}{r^{3\omega_{q}+1}}\Big).
\end{equation}

When a particle of energy $\omega$ is radiated from the black
hole, it transforms to
\begin{equation}\label{16}
    \dot{r}=\frac{1}{2}\Big(1-\frac{2(M-\omega)}{r}+\frac{Q^{2}}{r^{2}}-\frac{c}{r^{3\omega_{q}+1}}\Big).
\end{equation}

The imaginary part of the action is
\begin{equation}\label{17}
    ImS=Im\int P_{r}dr=Im\int\int dP_{r}dr=Im\int\int
    \frac{dH}{\dot{r}}dr,
\end{equation}
where we have used the Hamilton equation
$\frac{dH}{dP_r}=\dot{r}$, $H=M-\omega'\Rightarrow dH=-d\omega'$.
The imaginary part of the action takes then the form
\begin{equation}\label{18}
\begin{array}{lll}
    ImS&=&Im\int^{M-\omega}_{M}\int\frac{2dr}{1-\frac{2M}{r}+\frac{Q^{2}}{r^{2}}-\frac{c}{r^{3\omega_{q}+1}}}(-d\omega')\\
    &=&Im\int\frac{2\omega r^{2}dr}{(r-R_+)(r-R_-)}.
    \end{array}
\end{equation}

We used the tunnelling method of \cite{pw} to evaluate the
integral over $r$ and obtain
\begin{equation}\label{19}
    ImS=\frac{2\pi R_+^{2}}{R_+-R_-}\omega.
\end{equation}

Using the WKB approximation, the rate of radiation is expressed as

\begin{equation}\label{20}
    \Gamma\propto e^{-2ImS}=e^{-\beta\omega},
\end{equation}
where $\beta$ is the Boltzmann factor with inverse temperature
expressed as
\begin{equation}\label{21}
    \beta=\frac{1}{T}=\frac{4\pi
    r^2_+}{r_+-r_-}\Big(1+\frac{c(r^{-3\omega_{q}}_{+}(r_+-2r_-)-r^{1-3\omega_{q}}_{-})}{(r_+-r_-)^2}\Big).
\end{equation}

\begin{figure}[h!]
  \includegraphics[width=0.45\textwidth]{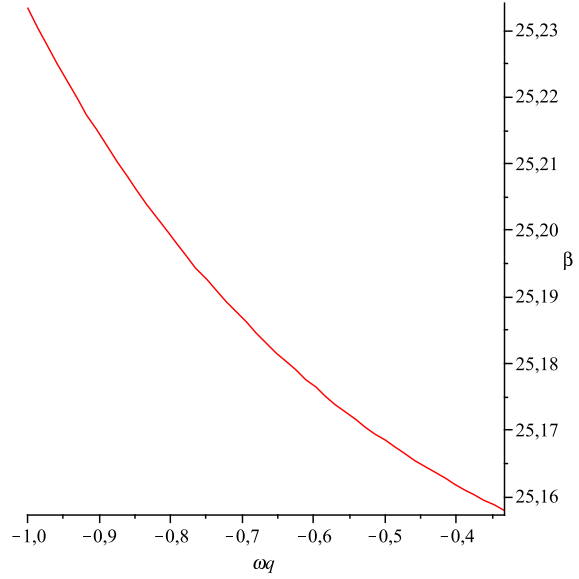}\\
  \caption{Variation of the Boltzmann factor with the state parameter of quintessence.}\label{fig5}
\end{figure}

Explicitly, we plot the variation of the Boltzmann factor with
respect to the state parameter of quintessence. Its behavior is
represented in Figure \ref{fig5}.

For a black hole with a charge such as Reissner-Nordstr\"om black
hole, the emitted particles can be charged. Thus, not only energy
conservation but also electric charge conservation should be
considered. Then, the radial null geodesics transforms to
\begin{equation}\label{22}
    \dot{r}=\frac{1}{2}\Big(1-\frac{2(M-\omega)}{r}+\frac{(Q-q)^{2}}{r^{2}}-\frac{c}{r^{3\omega_{q}+1}}\Big),
\end{equation}
where $q$ is the charge of the emitted particle.

The electromagnetic potential becomes
\begin{equation}\label{23}
    A_t=\frac{Q-q}{r}.
\end{equation}

The imaginary part of the action for the massive charged particle
is (\cite{jiang}):
\begin{equation}\label{24}
\begin{array}{lll}
    ImS&=&Im\int^{t_{f}}_{t_{i}}(L-P_{A_{t}}\dot{A_{t}})dt\\
    &=&Im\int^{r_{fe}}_{r_{ie}}(P_{r}\dot{r}-P_{A_{t}}\dot{A_{t}})\frac{dr}{\dot{r}}\\
    &=&Im\int^{r_{fe}}_{r_{ie}}\Big[\int^{(P_{r},P_{A_{t}})}_{(0,0)}(\dot{r}dP'_{r}-\dot{A_{t}}dP'_{A_{t}})\Big]\frac{dr}{\dot{r}},
\end{array}
\end{equation}
where $r_{ie}$ and $r_{fe}$ represent the localization of the
event horizon before and after the particle with energy $\omega$
and charge $q$ tunnels out. $\dot{r}$ and $\dot{A_t}$ are given by
the Hamilton's canonical equation of motion
\begin{equation}\label{25}
\left\{\begin{array}{ll}
    \dot{r}=\frac{dH}{dP_r}|_{(r;A_t,P_{A_t})},&dH|_{(r;A_t,P_{A_t})}=d(M-\omega)\\
    \dot{A_t}=\frac{dH}{dP_{A_t}}|_{(A_t;r,P_{r})},&dH|_{(A_t;r,P_{r})}=\frac{Q-q}{r}d(Q-q).\\
    \end{array}\right.
\end{equation}

Substituting equations (~\ref{22}) and (~\ref{25}) into
(~\ref{24}), we obtain
\begin{equation}\label{26}
\begin{tabular}{l}
    $ImS$=$Im\int^{\small{(M-\omega,Q-q)}}_{\small{(M,Q)}}[dH|_{(r;A_t,P_{A_t})}-dH|_{(A_t;r,P_{r})}]\frac{dr}{\dot{r}}$\\
    =$Im\int_{r_{ie}}^{r_{fe}}\int_{\small{(M,Q)}}^{\small{(M-\omega,Q-q)}}\frac{2[d(M-\omega')-\frac{Q-q'}{r}d(Q-q')]dr}{\Big(1-\frac{2(M-\omega')}{r}+\frac{(Q-q')^{2}}{r^{2}}-\frac{c}{r^{3\omega_{q}+1}}\Big)}.$
\end{tabular}
\end{equation}

Using the method of \cite{pw}, we can get
\begin{equation}\label{27}
    ImS=-Im\int^{r_{fe}}_{r_{ie}}(i\pi
    r)dr=\frac{\pi}{2}(r^{2}_{ie}-r^{2}_{fe}).
\end{equation}

Using the WKB approximation, we can get the tunnelling rate of
radiation
\begin{equation}\label{28}
    \Gamma\propto e^{-2ImS}=e^{\pi(r^{2}_{fe}-r^{2}_{ie})}=e^{\Delta
    S_{EH}},
\end{equation}
where $\Delta S_{EH}$ denotes the change of Bekenstein-Hawking
entropy at the even horizon before and after the particle
tunnelled out, expressed as
\begin{equation}\label{29}
    \Delta
    S_{EH}=\pi\Big(R^{2}_{+}(M-\omega,Q-q)-R^{2}_{+}(M,Q)\Big),
\end{equation}
where
\begin{equation}\label{30}
  R^{2}_{+}(x,y)=(x+\sqrt{x^{2}-y^{2}})^{2}+\frac{c(x+\sqrt{x^{2}-y^{2}})^{2-3\omega_{q}}}{\sqrt{x^{2}-y^{2}}}.
\end{equation}

The change of Bekenstein-Hawking entropy at the even horizon can
then be written as
\begin{equation}\label{31}
    \Delta
    S_{EH}=\Delta S_{0}+\Delta S_{q},
\end{equation}
where
\begin{equation}\label{32}
\begin{array}{lll}
    \Delta
    S_{0}&=&\pi\Big[(M-\omega+\sqrt{(M-\omega)^{2}-(Q-q)^{2}})^{2}\\
    &&\qquad-(M+\sqrt{(M)^{2}-(Q)^{2}})^{2}\Big],\\
    \Delta S_{q}&=&\pi
    c\Big[\frac{\Big(M-\omega+\sqrt{(M-\omega)^{2}-(Q-q)^{2}}\Big)^{2-3\omega_{q}}}{\sqrt{(M-\omega)^{2}-(Q-q)^{2}}}\\
    &&\qquad-\frac{\Big(M+\sqrt{(M)^{2}-(Q)^{2}}\Big)^{2-3\omega_{q}}}{\sqrt{(M)^{2}-(Q)^{2}}}\Big],
\end{array}
\end{equation}
$\Delta S_{0}$ is the free variation of the entropy and $\Delta
S_{q}$ is the contribution of the entropy variation due to the
quintessence.

\begin{figure}[h!]
  \includegraphics[width=0.45\textwidth]{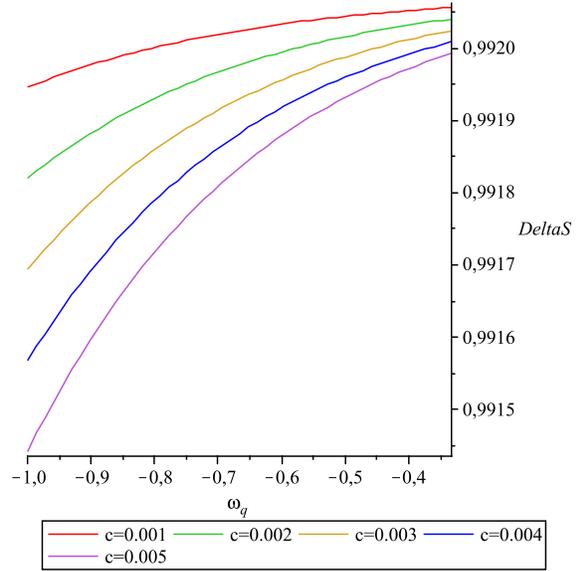}\\
  \caption{Variation of the entropy versus the state parameter of quintessence.}\label{fig:6}
\end{figure}

Supposing that the mass and charge of the black hole are uniformly
distributed and considering that the black hole radiates particles
with energy and charge proportional to the total mass and charge,
respectively, with the same coefficient of proportionality $a$,
\begin{equation}\label{33}
    \omega=aM,\qquad q=aQ, \qquad a<<1,
\end{equation}
the variation of entropy can be written as
\begin{equation}\label{34}
\begin{array}{lll}
    \Delta S&=&
   \Delta S_{0}+\Delta
   S_{q}\\
   &\simeq&-a\pi\Big[2(M+\sqrt{M^{2}-Q^{2}})^{2}\\
   &&~~\quad+(1-\omega_q)c\frac{\Big(M+\sqrt{M^{2}-Q^{2}}\Big)^{2-3\omega_{q}}}{\sqrt{M^{2}-Q^{2}}}\Big].
\end{array}
\end{equation}
Its behavior is plotted in Figure \ref{fig:6}.

Through this figure, we can see that the variation of entropy is
decreasing when decreasing $\omega_q$. We can also see that it is
decreasing when increasing $c$.

\section{Summary and Conclusion}
\label{conc} In summary, QNMs of a scalar field perturbation
around a Reissner-Nordstr\"om black hole were evaluated using the
third order WKB approximation. The results of table~\ref{tab:1}
are obtained without the presence of quintessence while those of
table~\ref{tab:2} are obtained under the presence of quintessence
for some values of the state parameter of the quintessence. The
Boltzmann factor with inverse temperature was also derived and its
behavior is plotted when varying the state parameter of
quintessence. The behavior of the variation of entropy is also
plotted when varying $c$ and $\omega_q$, respectively.

Through the above tables, we can remark that the absolute values
of the imaginary parts of the quasinormal frequencies under
quintessence are smaller compared to those without quintessence,
for fixed set of $l$ and $n$. Moreover, we can remark through
table 2 that these values decrease when decreasing $\omega_q$.
From the variation of the Boltzmann factor plotted bellow, we can
see that it is increasing when decreasing $\omega_q$. From the
behavior of the variation of entropy with respect to $c$ and
$\omega_q$, respectively, we can remark that this variation of
entropy is decreasing when increasing $c$ or when decreasing
$\omega_q$, denoting that the rate of radiation is decreasing.
Decreasing $\omega_q$ for fixed $c$, or increasing $c$ for fixed
$\omega_q$ means increasing the density of quintessence. Thus, we
can conclude that when increasing the density of quintessence
surrounding the Reissner-Nordstr\"om black hole, the QNMs damp
more slowly and the black hole radiates at slower rate.

\end{document}